**Waveguide Engineering of Graphene's Nonlinearity**

Kelvin J. A. Ooi, Lay Kee Ang and Dawn T. H. Tan*

Engineering Product Development, Singapore University of Technology and Design, 20 Dover Drive, Singapore 138682

*Corresponding author: dawn_tan@sutd.edu.sg

Abstract: Graphene has recently been shown to possess giant nonlinearity, however, the utility of this nonlinearity is limited due to high losses and small interaction volume. We show that by performing waveguide engineering to graphene's nonlinearity, we are able to dramatically increase the nonlinear parameter and decrease the switching optical power to sub-watt levels. Our design makes use of the hybrid plasmonic waveguide and careful manipulation of graphene's refractive index by tuning its Fermi level. The ability to tailor the nonlinear parameter in graphene based waveguides via the Fermi level provides a paradigm of nonlinear optics devices to be realized.

Graphene is a single sheet of carbon atoms with a honey-comb lattice structure, a naturally-occurring 2D material which has risen to popularity in many areas of research and technology [1]. In the field of photonics, graphene has enjoyed widespread research attention in various optical devices like photodetectors [2,3], modulators and switches [4-6], optical logic gates [7], and lasers [8,9]. Some of the strengths of graphene manifest in its unique optical properties, the most common include its large tunable refractive index, high confinement factor, and a universal absorption of 0.3% [10].

Of graphene's many unique optical properties, one of them is the giant nonlinearity which has been measured by several groups [11–13]. However, there are limitations to the utility of graphene's nonlinearity due to graphene's high losses and small interaction volume, as was pointed out by Khurgin in a recent letter [14]. In his analysis, using a normal-incidence configuration, graphene's nonlinear index $n_2$ could reach as high as $10^{-8}$ m$^2$/W, but over 1000 layers are required to perform a π-phase shift; otherwise, a dielectric waveguide configuration could be used but with a drastically lowered effective $n_2$ of $10^{-17}$ m$^2$/W. While Khurgin's analysis is valid to a certain extent, there is an omission to analyze graphene's nonlinear performance when incorporated into more sophisticated waveguide structures. In this Letter, we attempt a more rigorous theoretical analysis of several dielectric and plasmonic-based waveguide structure that could enhance the performance of graphene's nonlinear performance. We will base our analysis on the telecommunications wavelength of 1550nm, and consider only waveguide structures that are practical to fabricate. Further, the ability to tune graphene's Fermi level either through chemical doping or electric gating has led to new approaches for achieving broadband optical modulators [15,16] and polarizers [17,18]. This unique characteristic also has the potential to be applied to nonlinear optics applications. We show in this letter that waveguide structures integrated with graphene have nonlinear parameters which depend strongly on graphene's Fermi level. This ability to tune the nonlinear properties of a graphene based waveguide via electric gating or chemical doping could enable a paradigm of nonlinear optics applications to be realized.

The performance of waveguide structures is dependent on the interplay between the refractive indices of the waveguide's constituent materials. Hence, we will start by examining the refractive index of graphene, $n_{gr}$, which can be found from its 2D conductivity [19]

$$\sigma^{(1)}(\omega) = \frac{ie^2(2k_BT)}{\pi\hbar^2(\omega+i\nu_1)}\left\{\frac{E_F}{2k_BT} + \ln\left[2\exp\left(-\frac{|E_F|}{k_BT}\right)+1\right]\right\}$$
$$+ \frac{ie^2}{4\hbar}\left\{0.5 + \tan^{-1}\left[\frac{\hbar(\omega+i\nu_2)-2|E_F|}{2k_BT}\right] - \frac{i}{2\pi}\ln\left(\frac{[\hbar(\omega+i\nu_2)+2|E_F|]^2}{[\hbar(\omega+i\nu_2)-2|E_F|]^2+(2k_BT)^2}\right)\right\}$$
(1)

which is a function of the radian frequency ω, relaxation frequencies $\nu_1$ and $\nu_2$ (values taken from ref. [13]), and Fermi-level $E_F$ at room temperature T = 300K. Then, by taking the Drude expression

$$\varepsilon(\omega) = 1 + \frac{i\sigma^{(1)}(\omega)}{\varepsilon_0 \omega d_{eff}}$$
(2)

where $d_{eff}$ is the graphene's layer thickness approximated to 0.3nm, $n_{gr}$ could be found by simply taking the square-root of the permittivity expression. The refractive index of graphene as a function of the Fermi-level is plotted in Fig. 1(a). There is a range of Fermi-level in which graphene's real and imaginary refractive indices drops to very small values, as highlighted in the shaded region of Fig. 1(a). In this range, which was first predicted by Mikhailov and Ziegler to occur at $3\hbar\omega/5 < E_F < \hbar\omega/2$ [20], graphene is transparent but optical confinement and waveguiding is poor.

Next, we will examine the nonlinear refractive index of graphene, $n_{2,gr}$, by using Khurgin's formula $n_{2,gr} = \chi^{(1)}(\omega)/I_{sat}$ and $I_{sat} = 2(\omega E_F/ev_F)^2/\eta_0$ [14]. However, we modified Khurgin's original expression by using the susceptibility expression derived from Eq. (1) that also takes into account the interband contribution and temperature effects. Using this formulation, we find that $n_{2,gr}$ may take positive or negative values in specific ranges of Fermi-levels; nevertheless we will only consider the magnitude of the nonlinear refractive index, $|n_{2,gr}|$, as plotted in Fig. 1(b).

$|n_{2,gr}|$ does not completely tell us about graphene's nonlinearity. To characterize the effective nonlinearity, we would also need to take into account the effective mode area $A_{eff}$ of the waveguide through the nonlinear parameter [21]

$$\gamma = \frac{\omega}{c}\frac{\iint S_z^2 n_2(x,y)dxdy}{(\iint S_z dxdy)^2}$$
(3)

This parameter elucidates the importance of waveguide design to maximize the optical power inside the graphene layer. Further, we can define $P_\pi = \pi/(\gamma L_{eff,max})$ as the π-shift optical switching power, where the maximum achievable effective length $L_{eff,max}$ could be found from the damping factor α through $L_{eff,max} = 1/\alpha$ [22, 23].

We begin our analysis of a simple λ/2 rectangular waveguide – a 1550nm x 775nm waveguide core made of PMMA (n=1.47), surrounded by air, with a graphene layer embedded in the middle as shown in Fig. 2(a). From Fig. 2(b), this waveguide has an average $A_{eff}$ of ~1.44μm². The $A_{eff}$ is small when graphene exhibits high refractive indices at Fermi-levels below 0.4eV, indicating that weak slab-guiding occurs in the graphene layer. However, at this range of Fermi-levels, there is an opacity tradeoff as shown in Fig. 2(c). Meanwhile, the trend for γ showed in Fig. 2(d) follows closely to that of $|n_{2,gr}|$ in Fig. 1(b), ranging in the order of $10^{-1}$ to $10^1$ W$^{-1}$m$^{-1}$. We note that the γ values are obtained easily from $\gamma_{eff} = |n_{2,gr}|_{eff} / A_{eff}$, and by using the dilution principle from Khurgin to find $|n_{2,gr}|_{eff} = |n_{2,gr}| \times 2d_{eff} / \lambda$ [14]. The switching power is very large, predominantly in the order of $10^2$ – $10^4$ W as shown in Fig. 2(e), which is similar to Khurgin's prediction.

We now proceed to examine plasmonic-type waveguide structures, which have the ability to confine optical modes down to sub-wavelength dimensions [24]. We first examine the metal-insulator-metal (MIM) type. The waveguide width in the x-direction is maintained at 1550nm, while the dielectric layer height in the y-direction is reduced to 20 – 100nm, sandwiched in between two gold layers, as shown in Fig. 3(a). The gold refractive index is obtained from Palik's handbook [25]. γ depends on the optical localization in the nonlinear layer, which is a strong function of both refractive index and the dielectric layer height, as shown in Fig. 3(b). Unlike dielectric waveguides, optical power is confined better in a lower-index dielectric layer, and this occurs in the Fermi-level window of 0.45eV – 0.55eV. In this refractive index range, γ could exceed 600 W$^{-1}$m$^{-1}$ when the waveguide height is shrunk down to 20nm. We expect even higher γ values when the waveguide height is further reduced, but for device feasibility, optical coupling to such narrow waveguides might not be efficient. Correspondingly, the switching power is reduced to below 100W as shown in Fig. 3(c).

We now turn our attention to hybrid plasmonic waveguides (HPWs), which has seen widespread applications in optical waveguides and devices [26]. HPWs are formed by having a low-index dielectric layer sandwiched in between a metal layer and another high-index dielectric layer, which is usually a semiconductor like silicon [27,28]. The popularity of the HPW lies not only in its characteristic balance between low-loss and high confinement waveguiding, but also in its flexibility and compatibility for electrical, photonic and plasmonic integration [29]. Recently, it has been shown that the HPW is able to improve the nonlinearity of organic polymers [30].

In our nonlinear waveguide, the HPW consists of a gold-graphene-dielectric waveguide stack as shown in Fig. 4(a). To ensure high confinement inside the graphene layer, the refractive index contrast between graphene and the dielectric has to be high. Consequently, from Figs. 4(b) and 4(c) we observe the best nonlinear performance in both γ and switching power at the lowest and highest achievable refractive indices for the graphene and dielectric layer respectively. In comparison to dielectric-embedded and MIM embedded graphene nonlinear waveguides, there is an even greater performance improvement, where we see γ reaching the order of $10^5$ W$^{-1}$m$^{-1}$ from Fig. 4(b), while the switching power is lowered to below 10W from Fig. 4(c).

Here we shall evaluate the nonlinear performance of a practical graphene HPW as shown in Fig. 5(a). We use silicon (n=3.48) as our dielectric layer for advantages of its high refractive index and compatibility with most silicon photonics platform. The gold-graphene-silicon waveguide sits on top of a buried oxide layer, and the structure is encapsulated by PMMA. The graphene layer requires a Fermi-level of 0.51eV, and this is conveniently provided for since graphene is adsorbed on the gold surface [31]. The silicon material is also assumed to have a weak nonlinear coefficient of $6 \times 10^{-18}$ m$^2$/W [32]. Most of the nonlinear performance parameters are affected by the dimensions of the silicon layer. We find that the optimized γ lies in the dimension of around 260nm×260nm from Fig. 5(b), with high values in the order of $10^5$ W$^{-1}$m$^{-1}$. However, α increases and decreases linearly with layer width and height respectively from Fig. 5(c). Hence, the overall required switching power is found to be lowest at width×height of 210nm×420nm from Fig. 5(d), at a sub-watt level of 0.534W. Based on this switching power, we can also find the characteristic 3dB phase-shift bandwidth $\phi$, defined as $(I/I_0)_{3dB} = \cos^2(\phi/2)$, which occurs at the phase-shift of at least $\pi/2$. We find that the 3dB phase-shift bandwidth is very large, from 1.50 – 1.70 µm, as seen from Fig. 5(e).

Next, we would also like to make some performance comparison across materials for nonlinear waveguides, in particular the chalcogenides which has made breakthroughs in the recent years [33]. One of the contention is that graphene's nonlinear performance is only comparable, if not higher, than that of the chalcogenides. The main reason given was that graphene's high damping loss would eventually diminish its advantage of having a high nonlinear index [14]. While this is true for conventional dielectric waveguide structures as seen from the beginning of our analysis, we would like to point out that the HPW structure has significantly improved the nonlinear performance. Compared to chalcogenides from literature [33], the $\gamma$ for graphene HPW could be easily $10^2 - 10^5$ times higher, effectively mitigating the high losses from $\alpha$ which is only higher by $10^1 - 10^4$ times. For a direct quantitative comparison of nonlinear waveguides with the same configuration, if we consider the MIM geometry in Fig. 3(a) and replace graphene with a hypothetically 0.3nm thick $As_2S_3$ film with refractive index of 2.44, $n_2$ of $1 \times 10^{-16}$ m$^2$/W, and assuming no material loss, we could only obtain $\gamma$ less than 10 W$^{-1}$m$^{-1}$ and switching power exceeding 1kW from our simulation studies. Similarly, if we consider the hybrid waveguide geometry in Fig. 5(a) and replace graphene with the hypothetically 0.3nm thick $As_2S_3$ film, we evaluated a maximum $\gamma$ of only 555 W$^{-1}$m$^{-1}$ and minimum switching power of 15.3W.

In conclusion, we have shown that graphene's high nonlinearity could be fully exploited through engineering of the Fermi level and waveguide geometry. Using effective structures like the HPW, the nonlinear parameter, $\gamma$, could be engineered to very high values, while the switching power dramatically lowered to sub-watt levels, and a 3dB phase-shift bandwidth of 200nm could be achieved. The design of the nonlinear HPW involves manipulating the refractive index of graphene by tuning its Fermi-level, and also the refractive index of the dielectric, to get a high index-contrast between the two layers. Silicon is our choice of dielectric due to its compatibility with the majority of silicon photonic devices and fabrication feasibility. The results of our analysis shows that graphene's nonlinear performance could rival that of chalcogenides, and is a promising nonlinear material deserving of more research attention.

Acknowledgments

Support from the MOE ACRF Tier 2 research grant, SUTD-MIT International Design Center and the SUTD-ZJU collaborative research grant are gratefully acknowledged.

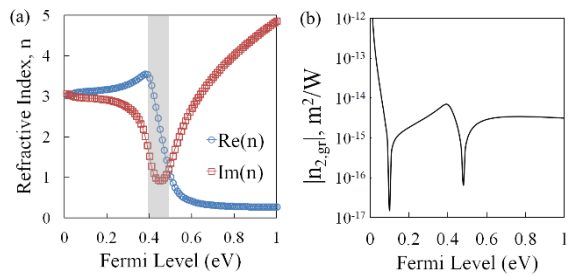

Figure 1 (a) Refractive index of graphene at 1550nm wavelength. (b) Nonlinear refractive index in $m^2/W$ of graphene at 1550nm wavelength.

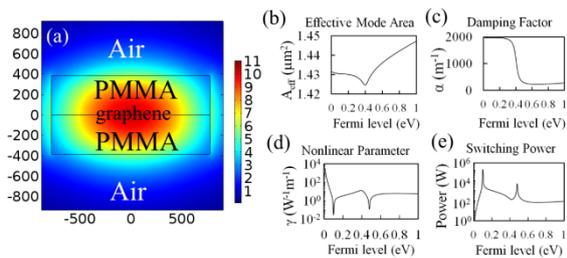

Figure 2 (a) Electric-field map of graphene embedded in PMMA waveguide. Performance parameters for the waveguide nonlinearity: (b) effective mode area, $A_{eff}$, (c) damping factor, $\alpha$, (d) nonlinear parameter, $\gamma$, and (e) switching power.

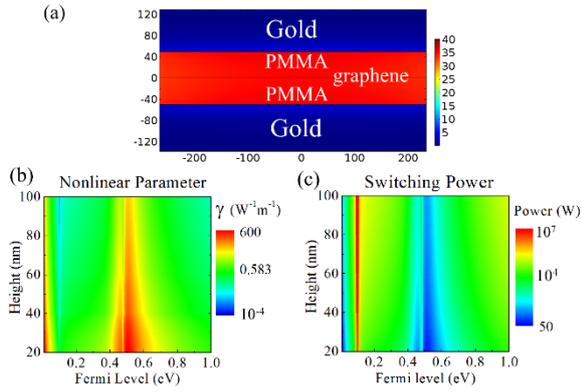

Figure 3 (a) Schematic and electric-field map for a gold–PMMA–graphene–PMMA–gold MIM-type plasmonic waveguide. Performance parameters for the waveguide nonlinearity as a function of dielectric layer height and graphene Fermi-level, plotted in logarithmic scale: (b) nonlinear parameter, γ, and (c) switching power.

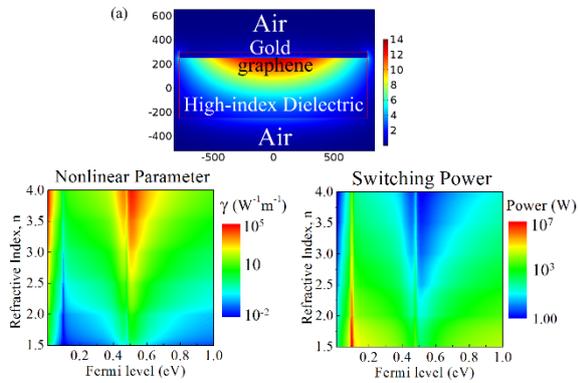

Figure 4 (a) Schematic and electric-field map for a gold-graphene-dielectric hybrid type plasmonic waveguide. Nonlinear performance parameters as a function of dielectric refractive index, n, and graphene Fermi-level, plotted in logarithmic scale: (b) nonlinear parameter, γ, and (c) switching power.

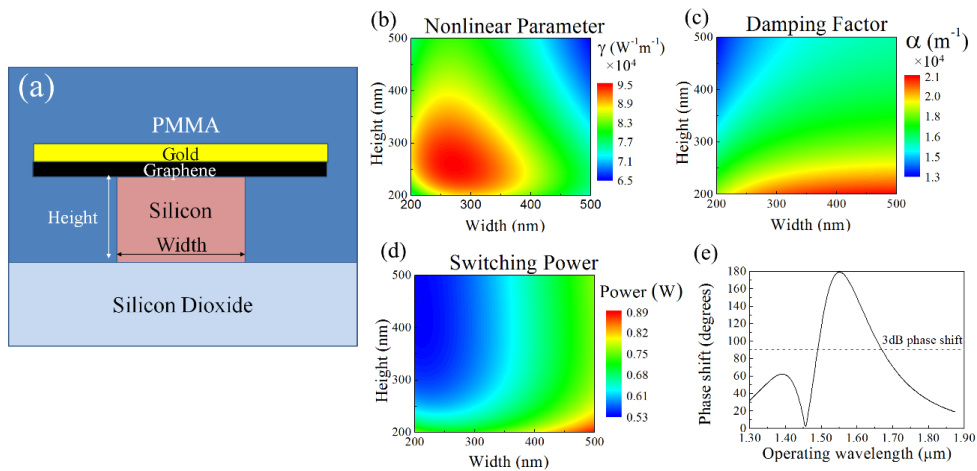

Figure 5 (a) Schematic structure of gold-graphene-silicon nonlinear waveguide, sitting on top of buried oxide and encapsulated by PMMA. Nonlinear performance parameters as a function of silicon height and width for (b) nonlinear parameter, γ, (c) damping factor, α, and (d) switching power. (e) The 3dB phase-shift bandwidth of the nonlinear waveguide.